# High-content stimulated Raman histology of human breast cancer

Hongli Ni,[1,#] Chinmayee Prabhu Dessai,[2,#] Haonan Lin,[1,#] Wei Wang,[3] Shaoxiong Chen,[4] Yuhao Yuan,[2] Xiaowei Ge,[1] Jianpeng Ao,[1] Nolan Vild,[1] and Ji-Xin Cheng[1,2]

[1]Department of Electrical and Computer Engineering, Boston University, 8 St. Mary's St., Boston, MA, 02215, USA

[2]Department of Biomedical Engineering, Boston University, 44 Cummington Mall, MA 02215, USA

[3]Hologic Inc., 250 campus drive, Marlborough, MA 01752

[4]Indiana University School of Medicine 340 West 10th Street, Fairbanks Hall, Suite 6200, IN 46202

[#]These authors contributed equally.

Corresponding author: jxcheng@bu.edu

## Abstract

Histological examination is crucial for cancer diagnosis, including hematoxylin and eosin (H&E) staining for mapping morphology and immunohistochemistry (IHC) staining for revealing chemical information. Recently developed two-color stimulated Raman histology could bypass the complex tissue processing to mimic H&E-like morphology. Yet, the underlying chemical features are not revealed, compromising the effectiveness of prognostic stratification. Here, we present a high-content stimulated Raman histology (HC-SRH) platform that provides both morphological and chemical information for cancer diagnosis based on un-stained breast tissues.

**Methods:** By utilizing both hyperspectral SRS imaging in the C-H vibration window and sparsity-penalized unmixing of overlapped spectral profiles, HC-SRH enabled high-content chemical mapping of saturated lipids, unsaturated lipids, cellular protein, extracellular matrix (ECM), and water. Spectral selective sampling was further implemented to boost the speed of HC-SRH. To show the potential for clinical use, HC-SRH using a compact fiber laser-based stimulated Raman microscope was demonstrated. Harnessing the wide and rapid tuning capability of the fiber laser, both C-H and fingerprint vibration windows were accessed.

**Results:** HC-SRH successfully mapped unsaturated lipids, cellular protein, extracellular matrix, saturated lipid, and water in breast tissue. With these five chemical maps, HC-SRH provided distinct contrast for tissue components including duct, stroma, fat cell, necrosis, and vessel. With selective spectral sampling, the speed of HC-SRH was improved by one order of magnitude. The fiber-laser-based HC-SRH produced the same image quality in the C-H window as the state-of-the-art solid laser. In the fingerprint window, nucleic acid and solid-state ester contrast was demonstrated.

**Conclusions**: HC-SRH provides both morphological and chemical information of tissue in a label-free manner. The chemical information detected is beyond the reach of traditional H&E and heralds the potential of HC-SRH for biomarker discovery.

# Introduction:

Breast cancer is one of the most prevalent cancers, accounting for ~15% of cancer-related deaths in women [1, 2]. As a highly heterogeneous disease, breast cancer requires detailed histopathology for diagnosis and treatment. Standard breast cancer histopathology includes Hematoxylin and Eosin (H&E) staining followed by Immunohistochemistry (IHC) staining, where H&E reveals the tissue morphology and IHC provides molecular information [3, 4]. However, these staining-based methods require labor-intensive and time-consuming sample preparation, which inhibits rapid diagnosis. Besides, the inter- and intra-lab differences in the staining protocol can induce result variation, leading to a diagnosis discordance [5, 6]. Staining differentiates tissue components based on their chemical contents, so tissue imaging techniques that require minimum sample preparation while providing rich chemical information are good candidates to overcome the above-mentioned problems.

As a label-free non-destructive chemical analysis approach, Raman spectroscopy has been widely applied in cancer research [7, 8]. Spontaneous Raman spectroscopy can hardly be applied to tissue histology due to its slow signal acquisition speed, typically several seconds per spectrum. Coherent Raman scattering microscopy, based on either coherent anti-Stokes Raman scattering (CARS) or stimulated Raman scattering (SRS), overcomes this limitation through coherent excitation, enabling high-speed chemical imaging [9-11]. Compared to CARS, SRS is more favorable for clinical practice because SRS is free of non-resonant background and can be performed under ambient light. SRS microscopy has been used for label-free cancer histology. By detecting lipid-rich and protein-rich regions with two SRS spectral channels, two-color stimulated Raman histology (SRH) was developed to visualize brain tumor morphology [12, 13]. The two-color SRH results correlate well with the H&E data, which enables rapid intraoperative brain tumor analysis when combined with deep learning algorithms [14, 15]. The two-color SRH was also successfully applied to other cancer types, including gastric, prostate, and breast cancer [16-18]. Besides two-color SRH, other studies utilized SRS to map microcalcifications which is a well-known breast lesion biomarker, for breast tissue classification [19, 20]. The two-color SRH and microcalcification-based approach both achieved high breast cancer detection accuracy when combined with machine learning models. Nevertheless, these methods only detect one or two chemical components, which compromises the chemical mapping capability of SRS. Recent studies show that spectroscopic SRS is capable of targeting multiple biomolecules of interest [21, 22]. As cancer development involves a variety of biomolecules, providing such rich chemical information is important for accurate subtyping of breast cancer [23, 24] and enables precise treatment.

Here, we demonstrated high-content stimulated Raman histology (HC-SRH) that exploits the rich chemical information in spectroscopic SRS. By implementing hyperspectral SRS imaging in the C-H spectral window and subsequent sparsity-penalized unmixing, we mapped 5 major chemical components in breast tissue: saturated lipids, unsaturated lipids, cellular protein, extracellular matrix (ECM), and water. The change in amount and relative ratio of these components were shown to be closely related to cancer progression [25-28]. In our work, successful high-content chemical mapping in the crowded C-H window was achieved by a least absolute shrinkage and selection operator (LASSO) regression algorithm for spectral unmixing [29, 30]. As imaging speed is a key parameter for clinical trials, we further boosted the speed of HC-SRH by one order of magnitude through spectrally selective sampling. Our HC-SRH results showed various tissue components for sub-typing of breast cancers. Moreover, using a multi-window fiber laser-based SRS microscope [31], we demonstrate compact HC-SRH for which the results agree well with the state-of-the-art SRS system. In the meanwhile, the broader spectral range of the multi-window fiber laser enabled HC-SRH to map more types of biomolecules. Our fingerprint results showcased a clear contrast for nucleic acid and solid-state ester as well as representative peaks for different amino acids. These contrasts are beyond the reach of C-H based SRH and show the potential of fingerprint SRH for discovering new biomarkers for cancer diagnosis [24].

## Methods and materials

### SRS imaging systems and data acquisition

**Figure 1a** shows the setup of a state-of-the-art SRS system based on a solid-state optical parametric oscillator (OPO). Without extra note, the SRS data was collected with this system. The solid-state OPO outputs two synchronized femtosecond laser pulse trains with 80 MHz repetition rate. The laser with shorter wavelength provides the pump beam and the other one fixed at 1040 nm provides the Stokes beam. For HC-SRH, the pump wavelength is centered at 800 nm. The Stokes beam is modulated by an acousto-optic modulator (AOM) at 2.4 MHz and pre-chirped by 15 cm SF57 rods. A motorized delay stage is put in the pump path to control the temporal overlapping of the two laser beams. The pump and Stokes are spatially combined with a dichroic mirror and then linearly chirped by 75 cm long SF57 glass rods. The combined beam is focused on the sample by a 60X water-immersion objective with 1.2 numerical aperture (NA). The transmission light is collected by an oil-immersion condenser with 1.4 NA and then filtered by a 1000 nm short pass filter to block the Stokes beam. The transmitted pump beam is detected by a photodiode. The SRS signal is conveyed as the modulation transfer from the Stokes beam to the pump beam. Extraction of the modulation is done with a lock-in amplifier. An SRS image is formed by laser scanning with a galvo mirror. A motorized sample stage is combined with the galvo mirror to cover a large field-of-view (FOV) for large-area tissue mapping. Hyperspectral SRS imaging is achieved through spectral focusing where the targeted spectral position is controlled by the temporal delay between the linearly chirped pump and Stokes laser pulses [21]. Second harmonic generation (SHG) imaging is performed in the same setup by replacing the photodiode with a photomultiplier tube. When performing SHG, only the Stokes laser is incident on the sample and a 520/70 bandpass filter is placed before the photomultiplier tube to filter out the Stokes beam.

**Figure 1b** presents the schematic of a compact SRS system based on a fiber OPO (FOPO) [32]. The setup of the FOPO-based SRS system is the same as the solid-state OPO-based system except for the modulation, hyperspectral scanning, and detection parts. The fiber laser outputs two synchronized picosecond laser pulse trains with 40 MHz repetition rate. The Stokes is modulated by a built-in modulator inside the laser at 6.7 MHz. This system realizes hyperspectral SRS by tuning the laser wavelength; therefore, no chirping rods and motorized stage are required. The FOPO can achieve arbitrary tuning between 700 and 3100 cm$^{-1}$ within 20 milliseconds. Because the fiber laser has high laser intensity noise, auto-balanced detection is implemented to improve signal-to-noise ratio. Briefly, two identical photodiodes detect the transmitted pump beam and a reference pump beam without passing the sample. The signal amplitudes from the two photodiodes are matched automatically by the variable gain amplifier and the proportional–integral–derivative controller. The laser noise is then suppressed by performing difference of the two amplitude-matched signals. The SRS signal is obtained by demodulating the differential signal with a lock-in amplifier.

The laser power incident on the sample is set as follows: pump 15 mW, Stokes 50 mW for solid-state OPO-based SRS; Stokes 50 mW for SHG; pump 15 mW, Stokes 100 mW for fiber laser-based SRS. All the SRS data is acquired with 10 μs pixel dwell time and 500 nm pixel size. The imaging time is around 1.0 minute for 1.0X1.0 mm$^2$ per spectral channel.

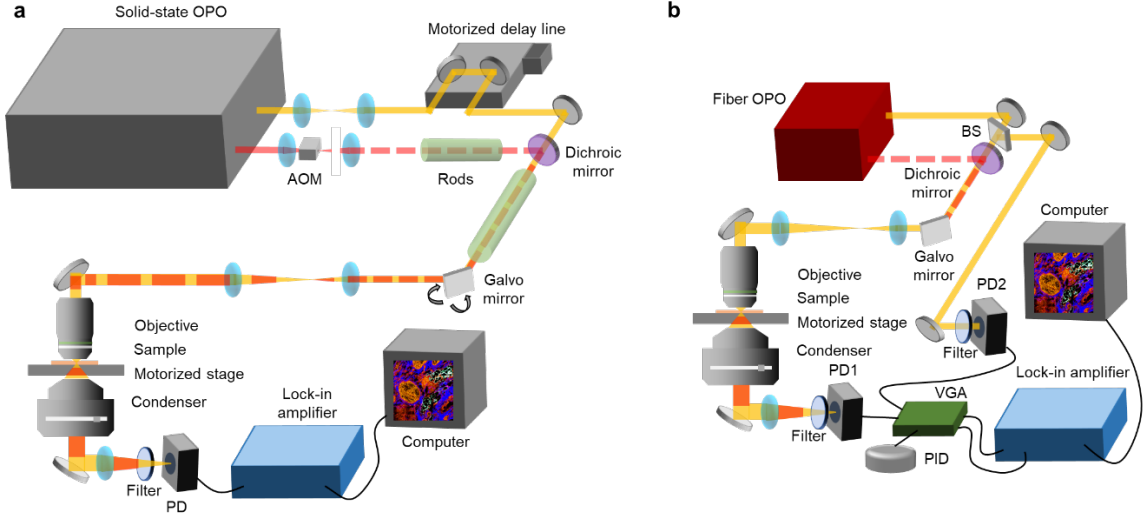

**Figure 1. System schematic.** (a) A state-of-the-art SRS imaging system with a solid-state laser. (b) A compact fiber laser-based SRS imaging system. OPO: optical parametric oscillator. AOM: acousto-optic modulator. PD: Photodiode. VGA: Variable gain amplifier. PID: proportional–integral–derivative controller. BS: Beam Splitter

**Sample preparation**

The breast cancer tissue samples are freshly frozen tissue sections purchased from Biochain Institute (Newark, CA). Each tissue section has a thickness around 5~10 µm and was mounted on a standard glass slide. For SRS and SHG imaging, the frozen tissue slide was washed using 1X phosphate-buffered saline (PBS) and then covered with a glass coverslip. Nail polish was used to seal the coverslip to prevent tissue dehydration. To acquire histology, a neighboring section of the tissue section used for SRS was prepared for standard H&E staining. The co-registration of HC-SRH and H&E was done manually.

**Data processing**

For every large-area SRS imaging data, the small-FOV SRS hyperspectral images were first fused into a large-FOV image with the ImageJ grid stitching toolbox. The fused hyperspectral image was then denoised by the spectral total variation algorithm [33]. Finally, the denoised SRS data was fed into the LASSO algorithm with reference spectra for generating concentration maps. The principle of LASSO can be expressed as:

$$\hat{C}_i = arg\min_{C_i}\left\{\frac{1}{2}||D(i,:) - C_i S||^2 + \beta||C_i||_1\right\}$$

Where $D(i,:)$ is the spectrum of the i[th] pixel in the data, $C_i$ represents the concentration for the targeted chemical components, $S$ is the reference spectra for the targeted chemical components. $\beta$ is the hyperparameter that can be fine-tuned to minimize channel crosstalk. We fixed $\beta$ =0.01 in this work. The reference spectra for unsaturated lipids, ECM, saturated lipids, and water were acquired from glycerol trioleate, breast tissue ECM, palmitic acid, and 1X PBS, respectively. The cellular protein spectrum is extracted by subtracting the unsaturated lipid spectrum from the breast cancer cell spectrum.

Spectral selective sampling is performed by a recursive feature elimination (RFE) algorithm. The principle of RFE is to iteratively discard the least contributing channel for the LASSO unmixing result. To determine the least important channel, every channel in the current spectrum will be tentatively discarded and a mean square error (MSE) will be calculated between the current LASSO result and the LASSO result generated with the full spectrum. The channel corresponding to the minimum MSE is the least important. The RFE algorithm stops when the number of current spectral channels reaches the targeted value. Because RFE has high computational complexity, we only fed a small data set of 2100 pixels to the algorithm to accelerate the selection process. The 2100 pixels consist of 5 subsets, where every subset has a dominant chemical component and was randomly selected from a representative FOV. This design can balance the contribution of every chemical component in the selection process. The data processing flow for the spectral selective sampled SRS data is the same as the hyperspectral data.

**Results**

**HC-SRH maps 5 chemical contents of breast cancer tissue**

**Figure 2** demonstrates that HC-SRH in the C-H window can successfully map 5 major chemical components in breast cancer tissue. **Figure 2a** shows the merged pseudo-color concentration map with yellow for unsaturated lipids, red for cellular protein, blue for ECM, cyan for saturated lipids, and grey for water. The separated concentration maps are shown in **Figure 2b**, where the water channel is contrast-enhanced to visualize the details better. The reference spectra used for LASSO unmixing are also shown in **Figure 2b**. The major spectral difference between the cellular protein and ECM is the overall spectral blue shift. This blue shift can come from the increased hydrogen bonds [34] or can be related to the fibrils formation [35].

To validate the concentration mapping given by HC-SRH, we compared the results of HC-SRH imaging with standard H&E staining on adjacent tissue section (**Figure 2c**). The unsaturated lipid is rich in the cell cytoplasm and fat cell residues, which is corroborated by the literature that breast cancer cells and adipose cells contain rich unsaturated lipids [36, 37]. The fat cells are seen as holes in HC-SRH because the tissue slicing procedure punctured the fat cells and only membrane-like residues remained. The cellular protein and ECM maps in HC-SRH agree well with the cell and ECM area in the H&E result. The fat cell residue area also shows a high saturated fat content in HC-SRH, which is consistent with that breast fat contains rich saturated lipids [36]. In addition, the solid-form saturated lipids are more likely to be preserved in tissue slicing. The water channel indicates that water is distributed across the whole FOV while the ECM contains slightly higher water content, and the fat residue area is lower in water content. This distribution makes sense because the main constituent of ECM, collagen, is hydrophilic, whereas lipid is hydrophobic. An enlarged view of one small FOV corresponding to the white box in Figure 2a is shown in **Figure 2d with spectra of five representative pixels** for cell nuclei, cell cytoplasm, ECM, fat cell residue and empty area.

We further compared HC-SRH with SHG which is commonly used for collagen imaging (**Figure 2e**). A cancer-adjacent blood vessel FOV is selected for verification because this FOV is rich in ECM including collagen and elastin. The H&E result is acquired from an adjacent section from the section used for SRS and SHG. The ECM channel of the HC-SRH correlates well with the SHG result. Yet, there are two main differences between the ECM channel of HC-SRH and the SHG. First, the intensity differs significantly for large and small fibers in SHG but appears more uniform in HC-SRH ECM channel. This difference may because SHG has a strong signal for type I and II collagen as well as highly aligned fibers but less sensitive to other types of collagens and small fibers [38]. The SRS signal is based on the protein concentration therefore provides better contrast for the small fibers. The second difference is that the blood vessel wall is visible in HC-SRH ECM channel but not in SHG. This difference is because the ECM of the blood vessel wall is rich in elastin [39], which can be detected by SRS but has a very low SHG signal.

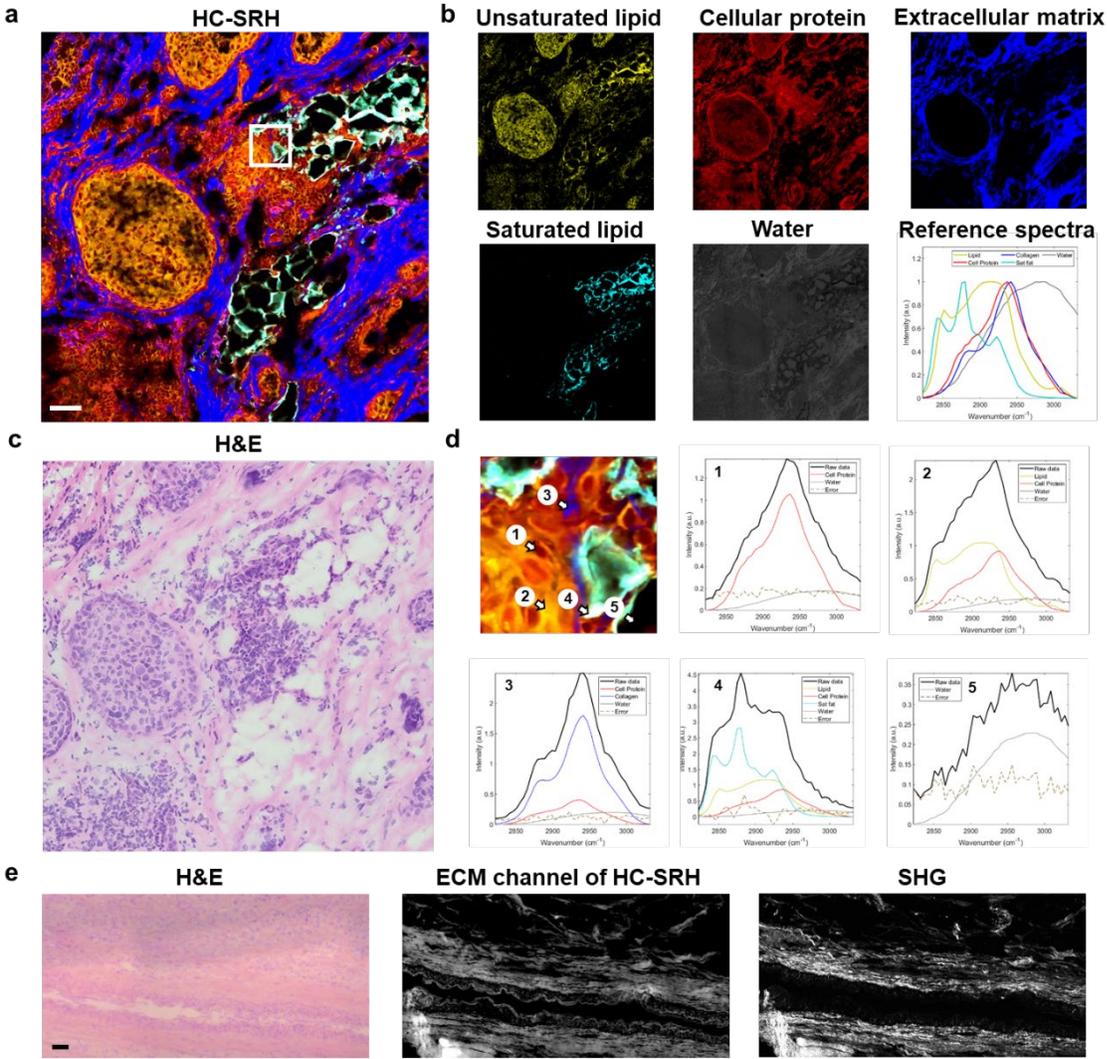

**Figure 2. HC-SRH of human breast cancer tissue slice.** (a) Merged concentration map for a breast cancer tissue sample. Yellow for unsaturated lipids, red for cellular protein, blue for ECM, cyan for saturated lipids, and grey for water. (b) Separate concentration maps for the 5 chemical components and the reference spectra corresponding to (a). (c) H&E result of a neighboring section to the tissue section used in (a). (d) Zoom-in view of the white box in (a) and SRS spectra of the five selected pixels as numbered. (e) Comparison of HC-SRH and SHG in mapping ECM in a cancer-adjacent vessel. Scale bar: 50 μm.

**Selective spectral sampling boosts HC-SRH speed by one order of magnitude**

The HC-SRH results shown in **Figure 2a** are generated with 45 SRS spectral channels covering 2820 to 3030 cm$^{-1}$, which required 40 minutes for the full field of view stack. As imaging speed is crucial for clinical practice, we boosted the speed of HC-SRH by reducing the number of spectral channels to be sampled. Theoretically, 5 chemical components require at least 5 channels to unmix. Therefore, we implemented the RFE algorithm to reduce the spectral channels to 5 while monitoring the HC-SRH result degradation to find a balance of between speed and image quality. **Figure 3** presents the HC-SRH result of two breast cancer tissue sections with 45, 20, 10, and 5 selected spectral channels. The HC-SRH quality is quantified by the structural similarity (SSIM) and peak signal-to-noise ratio (PSNR) to the 45-

channel result. The channel reduction does not lead to significant image quality degradation. Even with only 5 channels, the image quality is sufficiently enough for further analysis. Hence, the speed of HC-SRH can be boosted by 9 times by reducing 45 spectral channels to 5 selected ones (2837, 2880, 2908, 2951, and 3009cm$^{-1}$). When using different datasets for the RFE selection, the selected channels were slightly different but proved to have similar performance (Supplementary material section 1). Thus, there can be multiple good combinations for the selective sampling, which makes sense as the reference spectra in the C-H window cover a wide spectral range.

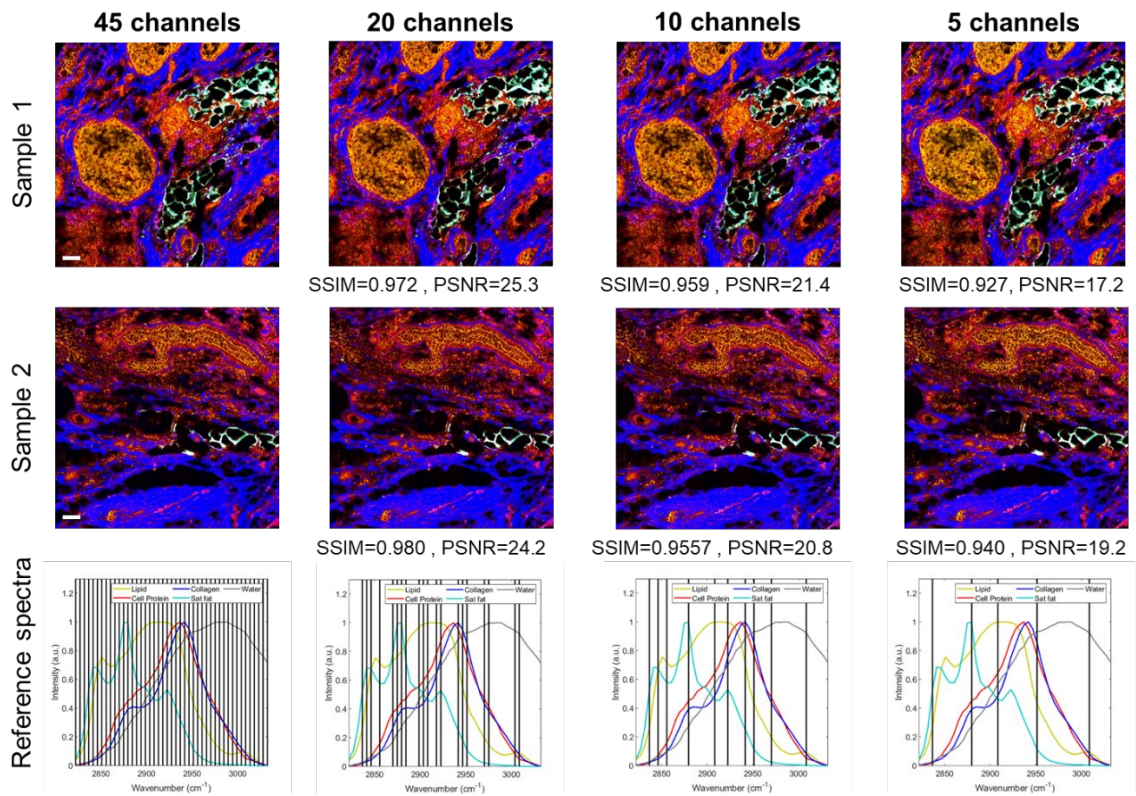

**Figure 3. High-speed HC-SRH via spectral selective sampling.** Scale bar: 50 μm. SSIM: structural similarity. PSNR: peak signal-to-noise ratio.

**HC-SRH provides excellent contrast for various breast tissue components**

With the 5 selected spectral channels, we demonstrated the histology capability of our method by imaging different types of breast lesion and structure, including typical ductal carcinoma in-situ (DCIS), usual ductal hyperplasia (UDH), normal duct, invasive ductal carcinoma (IDC), invasive lobular carcinoma (ILC), and chemotherapy lesion (Figure 4a-d). **Figure 4e-h** shows the H&E results of adjacent sections corresponding to **Figure 4a-d**. The SRH acquisition time for every tissue is less than 7 minutes.

We selected 10 regions of interest (ROI) from the four breast tissue sections to illustrate the performance of HC-SRH in detail. The ROI 1 is a typical DCIS where the cancer cells have grown inside the duct, and the ECM fibers nearby are stretched because of the over-growth. ROI 2 shows a necrosis area in the center of the DCIS duct, which is a signature of high-grade DCIS [40]. In HC-SRH, the necrosis shows an obviously higher cellular protein concentration than the surrounding cells, possibly due to cell condensation [41]. ROI 3 visualizes some stroma cells buried in the ECM, whose zoom-in view is shown in **Figure 4i**. ROI 4 shows IDC permeates the fat cell area, where the fat cell morphology differs from the healthy fat cells with smaller structures surrounded by more ECM, which is the signature of cancer-associated adipocytes. (**Figure 4j**). ROI 5 presents a normal duct structure. ROI 6 is a usual ductal hyperplasia structure whose zoom-in is **Figure 4k**. The duct cells in **Figure 4k** are overgrown, but the two-layer structure of the duct is still visible. ROI 7 is a blood vessel structure where the elastin around the vessel wall is shown as a mixture of ECM and cellular protein in HC-SRH. ROI 8 shows more intact fat cells close to the cancerous area. ROI 9 presents an ILC-like characteristic where cancer cells are aligned in a line shape (**Figure 4l**). ROI 10 shows a post-chemotherapy lesion consisting of newly grown collagen. The newly grown collagen is shown as a mixture of ECM and cellular protein in the HC-SRH, likely because the newly synthesized collagen is at an intermediate between fully grown collagen and cellular protein biomolecules. In summary, **Figure 4** indicates that HC-SRH has excellent contrast for duct, necrosis, stroma, blood vessel, fat cell, chemotherapy lesion, and epithelial cell morphology, which corresponds well with the H&E result. In addition, HC-SRH overperforms H&E in fat cell identification and distinguishing changes in the ECM composition.

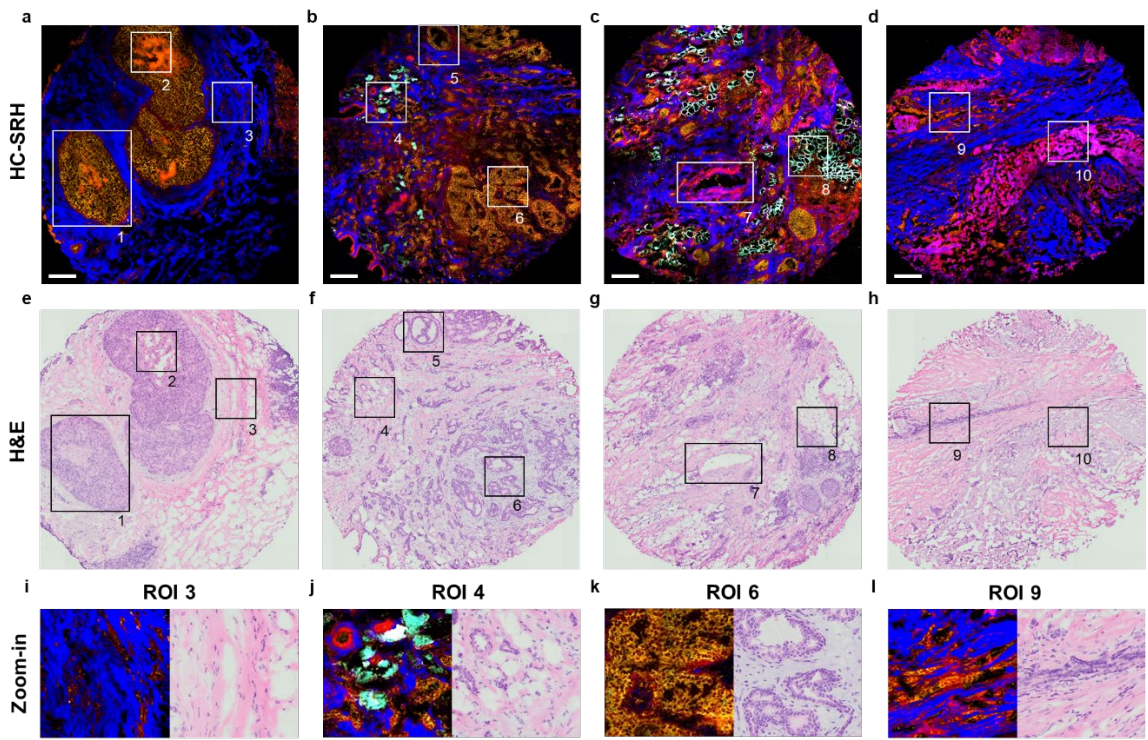

**Figure 4. Breast tissue features revealed by HC-SRH.** (a-d) HC-SRH of breast lesion tissue with 5-channel selective spectral sampling. a: typical ductal carcinoma in-situ (DCIS); b: mixture of usual ductal hyperplasia (UDH), normal duct, and invasive ductal carcinoma (IDC); c: IDC and invasive lobular carcinoma mixture (ILC); d: contains and chemotherapy lesion and some ILC residue (e-h) H&E of neighboring tissue section corresponding to (a-d). (i-l) Zoom-in of ROIs 3,4,6,9. Scale bar: 200 μm.

### HC-SRH with a compact FOPO laser

The results in **Figures 2-4** were acquired with a state-of-the-art SRS system operated with a solid-state OPO. The solid-state OPO is bulky and environment-sensitive, which is not suitable for clinical use. Furthermore, the complex optical alignment, e.g., spectral focusing, to achieve hyperspectral scanning requires well-trained users. A rapid-tuning picosecond fiber laser facilitates the system operation and is suitable for clinical translation. We performed HC-SRH on a recently built FOPO-laser-based SRS system [31]. **Figure 5** compares the HC-SRH result acquired with the FOPO-laser-based system and that with the solid-state OPO-based system. Comparing **Figure 5a** and **5b**, we saw that selective spectral sampling can still preserve most of the HC-SRH quality using the same selected wavenumbers as in section 3.2. Equivalent image quality was achieved with fiber laser-based system as with solid-state OPO-based system (Figure 5b and 5c).

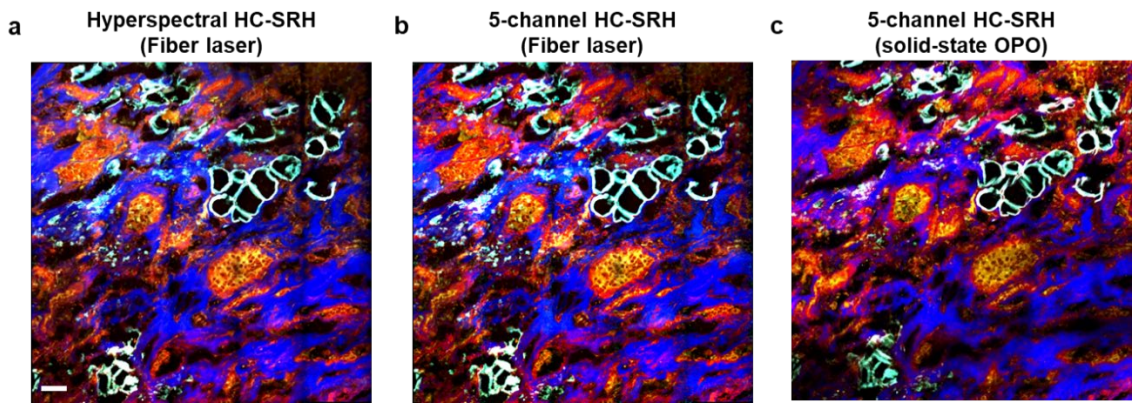

**Figure 5. HC-SRH with a compact fiber laser.** (a) Hyperspectral HC-SRH (61 spectral channel) acquired by the FOPO-laser-based SRS system. (b) Selective sampled HC-SRH (5 spectral channel) acquired by the FOPO-laser-based SRS system. The spectral channel position is the same with Figure 3. (c) Selectively sampled HC-SRH (5 spectral channel) acquired in solid state OPO-based SRS system. Scale bar: 50 μm.

**SRS imaging of breast cancer tissue in the fingerprint window**

Besides being clinically compatible, the FOPO-laser-based SRS system can complete fast tuning between two arbitrary wavenumbers within 20 ms. This rapid spectral tuning capacity enables convenient spectroscopic SRS imaging across the whole biologically relevant Raman window. Fingerprint and C-H SRS hyperspectral images of a breast cancer tissue sample are shown in Figure 6. With the C-H HC-SRH (**Figure 6a**), we distinguished between different tissue components as introduced in section 3.1. The fingerprint images at the 6 characteristic Raman peaks are presented in **Figure 6b**. The 790 $cm^{-1}$ and 1495 $cm^{-1}$ images corresponding to nucleic acid peaks can clearly visualize the cancer cell nuclei, which is beyond the reach of C-H SRH. Phenylalanine and carotenoids have a characteristic peak at 1005 $cm^{-1}$. Images at this wavenumber can help reveal the detailed composition of protein and lipids. The image at 1250 $cm^{-1}$ corresponding to the collagen further validates our ECM mapping in the C-H HC-SRH. Combining the 1300 $cm^{-1}$ image for lipids and the 1745 $cm^{-1}$ image for ester, we find that the lipids in the fat cell residue are highly esterized, but the lipids in cancer cell cytoplasm are not. Thus, combining information from the fingerprint region data and the C-H region data can provide valuable information for more precise cancer diagnosis.

The fingerprint spectra of the 4 ROIs in **Figure 6a** are shown in **Figure 6c**. The 4 ROIs are selected to represent cancer cell cytoplasm, cancer cell nuclei, ECM, and fat cells. As expected, the cytoplasm spectrum shows a series of characteristic peaks for lipids and protein; the cell nuclei spectrum has many peaks corresponding to the nucleic acid; ECM spectra show many collagen features, and fat cell residue spectra are dominated by lipid characteristics. Another point to be noted is that we see a split ester peak in 1730 and 1745 $cm^{-1}$ in the fat cell spectrum. This characteristic corresponds to a feature of solid-state fat [42], which is because the solid-state fat is better preserved than other lipids after tissue slicing. A detailed peak registration can be found in the supplementary material **Table T1**.

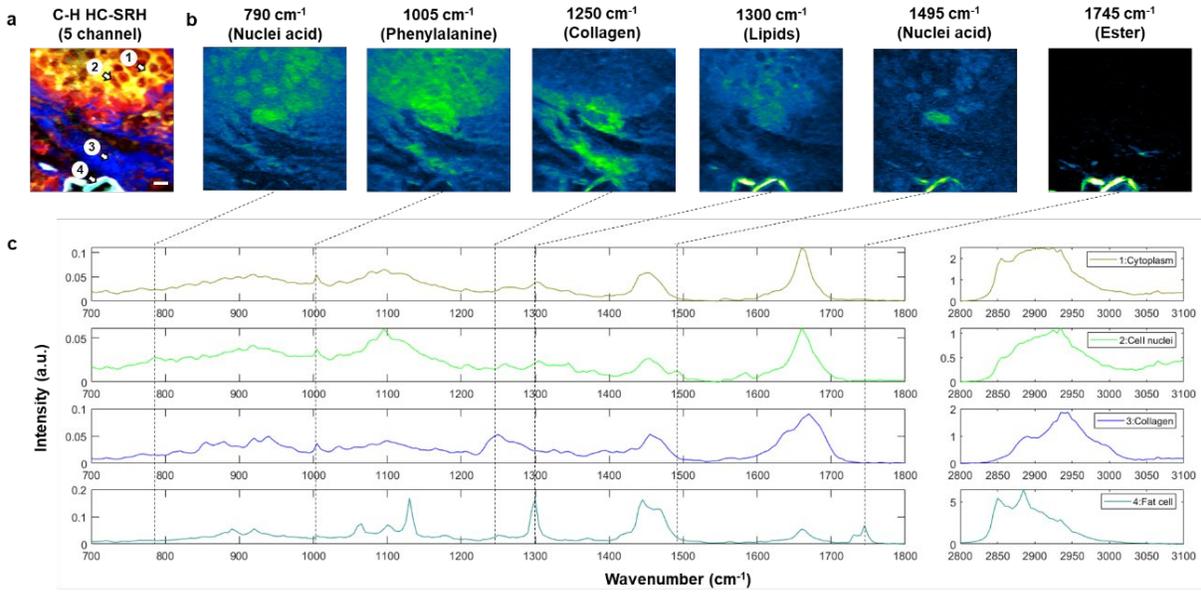

**Figure 6. Fingerprint SRS imaging of breast cancer tissue.** (a) C-H HC-SRH result. (b) Fingerprint SRS result at representative spectral position in the same FOV of (a). (c) Multi-window SRS spectra of the 4 selected ROIs in (a). Scale bar: 10 μm.

## Discussion

In this study, we have developed stain-free HC-SRH that provides breast cancer histology with rich chemical information. Compared to the well-developed two-color SRH, HC-SRH provides extra chemical information that can benefit cancer diagnosis in two aspects. First, HC-SRH can improve cancer classification accuracy which is of great importance in clinics. Machine learning has been widely applied to histology results to obtain diagnostic information automatically. Although decent classification accuracy can be obtained from tissue morphology, studies have indicated that chemical information can further improve classification performance [43]. For clinical application, classification accuracy is of great importance, and every improvement is worthwhile. Secondly, chemical information gained from HC-SRH can contribute greatly to our knowledge about cancer biology which can accelerate the discovery of new biomarkers. Other than relying on the morphological-based machine learning classification, HC-SRH can deepen our understanding of subtle chemical changes involved in cancer progression.

To further advance the HC-SRH method toward clinical trial, HC-SRH needs to be carried out on a large set of fresh biopsy samples to validate the technique's robustness and determine targeted chemical information. Imaging on fresh biopsy samples will avoid artifacts created by sample preparation. For example, during the slicing process of sample preparation, the fat cells were lost therefore appeared dark in our imaging results. Additionally, the water channel of HC-SRH was less informative because the externally added PBS solution largely contributes to the water in the current sample rather than the tissue itself. Previous study has shown that two-color SRH can be performed on fresh breast needle biopsy samples [17]. Therefore, HC-SRH can also be performed on the fresh samples. In this study, we did not establish chemical biomarkers for cancer diagnosis due to the limited number of samples. The informative fingerprint window has great potential for identifying cancer biomarkers [7], which requires a large dataset to validate. Once the chemical components to target are determined, spectral selective sampling can be performed to target the key components specifically. For example, microcalcifications [19, 20] and carotenoids [44] are indicated as biomarkers for breast cancer development, and both have characteristic peaks in the fingerprint window. In this work, we did not perform mapping of these two components due to the limited breast tissue samples, in which we did not observe obvious microcalcification or carotenoid content.

## Conclusion

We have developed a new approach for label-free cancer histology called high-content stimulated Raman histology (HC-SRH). The HC-SRH in the C-H window allows the mapping of major chemical contents in the breast tissue and provides excellent contrast for various tissue components. Through implementing spectral selective sampling, we boosted the speed of HC-SRH by ~1 order without sacrificing image quality. We also demonstrated that the HC-SRH is robust with a clinical-compatible fiber laser system. With the rapid, widely tuning capability of the FOPO, we extended the spectral coverage of the HC-SRH to the fingerprint window, which provides extra contrast for nucleic acid, amino acid, and solid-state fat in breast tissue. Collectively, these results show that HC-SRH is a promising tool for label-free cancer histology with rich chemical information.

## Acknowledgements

This project is supported by funds provided by Hologic to Boston University. We thank Dr. Linda Han for the helpful discussion on breast cancer diagnosis.

## Abbreviations

H&E: hematoxylin and eosin; IHC: immunohistochemistry; HC-SRH: high-content stimulated Raman histology; CARS: coherent anti-Stokes Raman scattering; SRS: stimulated Raman scattering; SRH: stimulated Raman histology; ECM: extracellular matrix ; LASSO: least absolute shrinkage and selection operator; OPO: optical parametric oscillator; AOM: acousto-optic modulator; NA: numerical aperture; FOV: field-of-view; SHG: second harmonic generation; FOPO: fiber optical parametric oscillator; PBS: phosphate-buffered saline; RFE: recursive feature elimination; MSE: mean square error; SSIM: structural similarity; PSNR: peak signal-to-noise ratio; DCIS: ductal carcinoma in-situ; UDH: usual ductal hyperplasia; IDC: invasive ductal carcinoma; ILC: invasive lobular carcinoma; ROI: regions of interest.

## Competing interests

The authors have declared that no competing interest exists.